\documentclass{revtex4}
\usepackage{amsmath}
\usepackage{slashed}
\usepackage{multirow}
\usepackage{graphicx,subfigure}
\usepackage{titlesec}

\begin{document}

\title{Covariant ChPT calculation of the hyperon forward spin polarizability}

\author{Astrid Hiller Blin}
\affiliation{Departamento de F{\'i}sica 
  Te{\'o}rica, Universidad de Valencia and IFIC, 
  Centro Mixto Universidad de Valencia-CSIC, Institutos 
  de Investigaci{\'o}n de Paterna, Aptdo. 22085, 46071 Valencia, 
  Spain} 

\author{Thomas Gutsche}
\affiliation{
  Institut f\"ur Theoretische Physik, Universit\"at T\"ubingen,
  Kepler Center for Astro and Particle Physics, 
  Auf der Morgenstelle 14, D-72076, T\"ubingen, Germany}

\author{Tim Ledwig}
\affiliation{Departamento de F{\'i}sica 
  Te{\'o}rica, Universidad de Valencia and IFIC, 
  Centro Mixto Universidad de Valencia-CSIC, Institutos 
  de Investigaci{\'o}n de Paterna, Aptdo. 22085, 46071 Valencia, 
  Spain} 

\author{Valery E. Lyubovitskij}
\affiliation{
  Institut f\"ur Theoretische Physik, Universit\"at T\"ubingen,
  Kepler Center for Astro and Particle Physics, 
  Auf der Morgenstelle 14, D-72076, T\"ubingen, Germany}
\affiliation{ 
  Department of Physics, Tomsk State University,  
  634050 Tomsk, Russia} 
\affiliation{Mathematical Physics Department, 
  Tomsk Polytechnic University, 
  Lenin Avenue 30, 634050 Tomsk, Russia} 

\begin{abstract}
We predict the values for baryon forward spin polarizabilities in fully covariant ChPT and including the virtual contributions of the 
spin-$3/2$ states. As the nucleon results are in good agreement with the experimental data and they do not depend 
on renormalization schemes, we extend the calculations to the hyperon sector.
\end{abstract}

\maketitle

\section{Introduction}
The forward spin polarizability $\gamma_0$ represents the deformation of a non-elementary particle 
relatively to its spin axis when submitted to Compton scattering of photons in the extreme forward 
direction. Once the scattering cross sections are experimentally obtained, it is connected to them via the sum rule~\cite{GellMann:1954db,TRHT98} 
\begin{equation}
  \gamma_0=-\frac1{4\pi^2}\int_{\omega_0}^{\infty}\mathrm{d}\omega
  \frac{\sigma_{3/2}(\omega)-\sigma_{1/2}(\omega)}{\omega^3},
\end{equation}
where $\omega$ is the photon energy and $\sigma_{3/2}$ and $\sigma_{1/2}$ the photo-absorption cross sections of parallel and antiparallel target 
and photon helicities, respectively.
Experimental values for the nucleons' $\gamma_0$ have been extracted in~\cite{Sandorfi:1994ku,Hildebrandt:2003fm,Pasquini:2010zr}, while 
theoretically it is determined from the spin-dependent piece of the Compton-scattering amplitude via~\cite{Bernard:1995dp,TRHT98}
\begin{equation}
  \gamma_0\left[\vec{\sigma}\cdot(\vec{\epsilon}
    \times\vec{\epsilon}~^\ast)\right]=
 - \frac{\mathrm{i}}{4\pi}\frac{\partial}{\partial \omega^2}
  \frac{\epsilon^\mu \mathcal{M}^{\text{SD}} _{\mu\nu}\epsilon^{*\nu}}
{\omega}\bigg|_{\omega=0}\\.
  \label{eqgamma0}
\end{equation}
The photon energies considered here probe the non-perturbative regime of QCD. Therefore it is reasonable to 
extract the amplitudes with the help of chiral perturbation theory (ChPT). The leading-order contributions to $\gamma_0$ 
appear at chiral order $p^3$, where the result depends only on well-known 
low-energy constants. Furthermore, the relevant pieces of the Compton-scattering amplitude for $\gamma_0$ 
show no divergences or power-counting breaking terms. Therefore, the results obtained are pure predictions. 

The nucleons' $\gamma_0$ has been thoroughly studied in~\cite{Bernard:1992qa,Lensky:2009uv,Lensky:2012ag,Lensky:2014efa}. It was shown in~\cite{Kao:2002cp,Bernard:2012hb,Lensky:2014dda} that the inclusion of the $\Delta(1232)$ resonance 
is important to better reproduce the experimental results. We extend these studies to the SU(3) sector, as has been done in the heavy-baryon 
formulation in~\cite{VijayaKumar:2011uw}, later improved in~\cite{Astrid_Diplom}. A study comparing different field-theoretical models is 
performed in~\cite{Holstein:2013kia}.

The SU(3) flavour version 
allows the inclusion of additional virtual-state contributions to the nucleon polarizability by extending the isospin-$1$ triplet of pions to 
the meson octet. Furthermore, it enables the prediction of the hyperon polarizabilities, the baryons of the isospin-$1/2$ octet with 
non-vanishing strangeness. 
Our calculations were 
performed in the frame of fully covariant ChPT, where we extended the model such as to include the isospin-$3/2$ decuplet of spin-$3/2$ resonances like the 
$\Delta(1232)$~\cite{Blin:2015era}.

\section{Treatment of Compton scattering in ChPT}
The relevant pieces of the SU(3) Lagrangian involving mesons, photons and baryons --- both of spin 1/2 and 3/2 ---  needed for the calculations in this work are the following:
\begin{align}
\nonumber  \mathcal{L}_{\phi B\Delta}
  =&\frac{F_0^2}4\text{Tr}\left(\nabla_\mu U\nabla^\mu U^\dagger+\chi_+\right)
+\text{Tr}\left(\bar{B}(\mathrm{i}\slashed{\mathrm{D}}-m)B\right)
+ \frac 12\text{Tr}\left(\bar{B}\gamma^\mu\gamma_5\left(D\left\{u_\mu,B\right\}+F\left[u_\mu,B\right]\right)\right)\\
&-\left(\mathrm{i}\bar{B}^{ab}\varepsilon^{cda}\left( \frac{\sqrt{2}\mathcal{C}}{F_0M_\Delta}
  \gamma^{\mu\nu\lambda}
    (\mathrm{D}_\lambda\phi)^{ce}
  +\frac{3eg_M}
          {\sqrt2m(m+M_\Delta)}
          Q^{ce}\tilde{F}^ {\mu\nu} 
          \right)(\partial_\mu\Delta_\nu)^{dbe}+ \text{H.c.}\right), 
  \label{eqlag}
\end{align}%
where the definitions of the constants and the different components are given in~\cite{SW79,Gasser:1987rb,Geng:2009hh,Geng:2009ys,Ledwig:2014rfa}. 
We added the missing couplings by extending the SU(2) Lagrangian from~\cite{Pascalutsa:2005ts,Pascalutsa:2005vq,Pascalutsa:2006up} to SU(3).  
We estimate the value for $g_M$ in an analogous way to the one described in~\cite{Bernard:2012hb}: 
We calculate the width 
of the electromagnetic decay of the $\Delta(1232)$,
\begin{align}\label{EqGM}
\Gamma_{\Delta}^\text{EM}=-2\text{Im}(\Sigma_{\Delta}^\text{EM})=
\frac{e^2g_M^2(M_\Delta-m)^3(M_\Delta+m)^3}{4M_\Delta^3m^2\pi},
\end{align} 
where $\Sigma_{\Delta}^\text{EM}$ is the electromagnetic $\Delta(1232)$ self-energy amplitude. The data on the hyperon electromagnetic decays 
is sparse, for which reason we opt not to do this calculation in SU(3). We obtain the value $g_M=3.16\pm0.16$, which leads to an important uncertainty in the results.

We follow the chiral counting scheme in~\cite{Pascalutsa:2002pi}, the 
so-called $\delta$ counting, where the small quantity $\delta=M_\Delta-m_N$ is treated as being of $\mathcal{O}(p^{1/2})$, and we 
introduce the couplings in a consistent dynamic, like in~\cite{Pascalutsa:1998pw,Pascalutsa:1999zz,Pascalutsa:2000kd}. 
The counting is a reasonable approximation when treating energies sufficiently far from the $\Delta(1232)$ mass, leading to the 
nominal order of a diagram being given by
\begin{equation}
  N=4N_L + \sum_{d=1}^\infty{d N_{d}} - 2P_\phi-P_B-\frac{1}{2}P_\Delta.
\end{equation}
In this case, the tree-level diagram with a virtual decuplet intermediate state is of $\mathcal{O}(p^{7/2})$. It is expected to give a dominant contribution. It is also exactly the diagram that is proportional to $g_M^2$, which makes our results very sensitive to this constant. Exception are the hyperons $\Sigma^-$ and $\Xi^-$, whose photon transition to members of the decuplet is forbidden. Therefore their results are mainly described by octet intermediate states, including only small corrections coming from the spin-3/2 states.

The Gauge is an issue which should be addressed at this point. The minimal coupling of the photon to the decuplet includes higher-order terms if 
one wants to use the fully covariant derivative $\mathrm{D}_\mu\Delta_\nu$. If one opts to have consistent power counting these higher-order terms should be neglected, but then gauge invariance is broken. Therefore special care has to be taken here. The solution we follow is analogous to the one proposed in~\cite{Lensky:2009uv} for the proton, and we extend it to the hyperon sector. In fact, this difficulty appears only for the charged octet-baryon members. The approach is to calculate two sets of diagrams separately: the one-particle reducible and the one-particle irreducible diagrams. The latter can be calculated in the usual way, summing over all possible isospin channels. As for the one-particle reducible diagrams, they are at first only 
computed for the charged meson channels. The reason for this is that in a world with only charged mesons, gauge invariance would be restored. 
The other channels' isospin factors are then chosen such that the ratio between the isospins of the two sets of diagrams is the same as in the 
case of charged mesons only. With this method, the gauge invariance is automatically obtained and the corrections that would come from 
taking the original isospin factors is of higher order.

\section{Results and discussion}

\begin{figure}
  \begin{center}
      \includegraphics[width=0.3\textwidth]{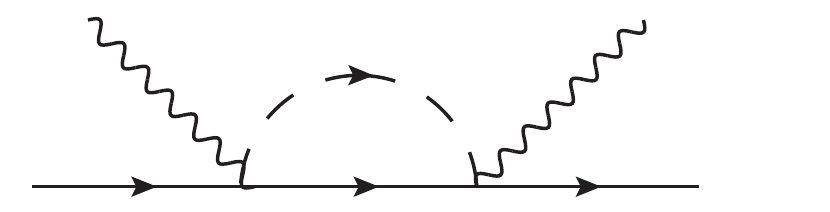}
      \includegraphics[width=0.3\textwidth]{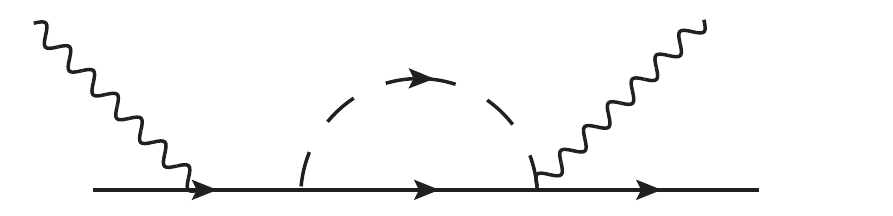}
      \includegraphics[width=0.3\textwidth]{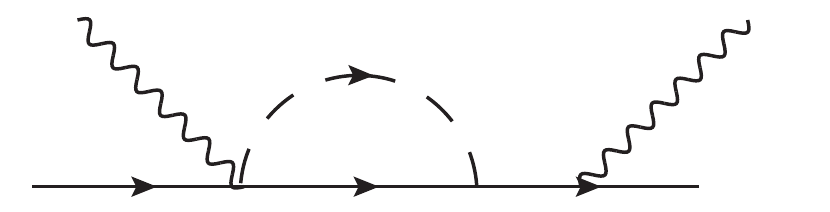}\\
      \includegraphics[width=0.3\textwidth]{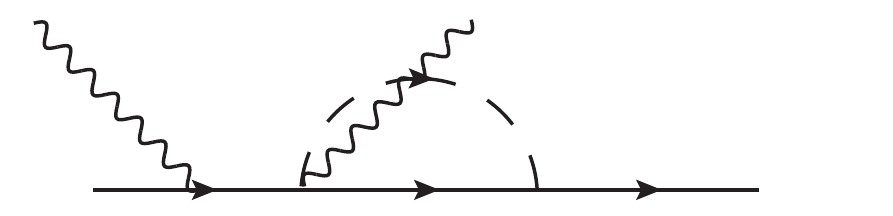}
      \includegraphics[width=0.3\textwidth]{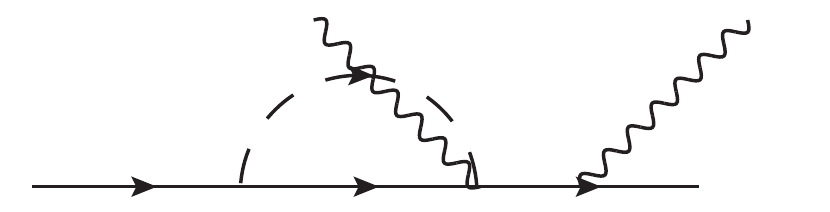}
      \includegraphics[width=0.3\textwidth]{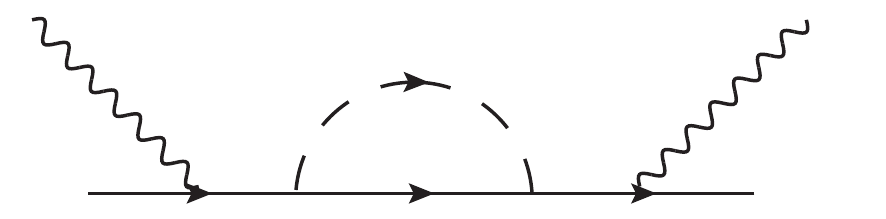}\\
      \includegraphics[width=0.3\textwidth]{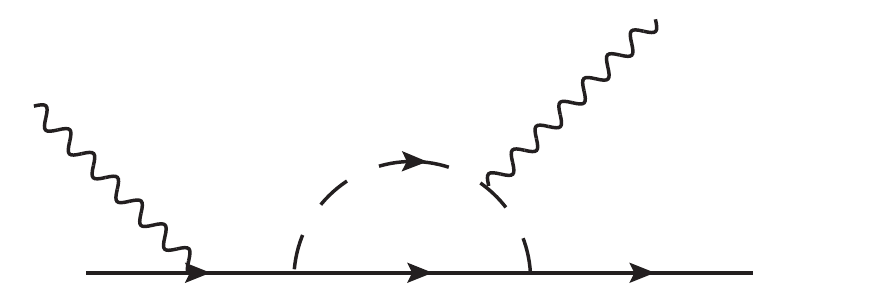}
      \includegraphics[width=0.3\textwidth]{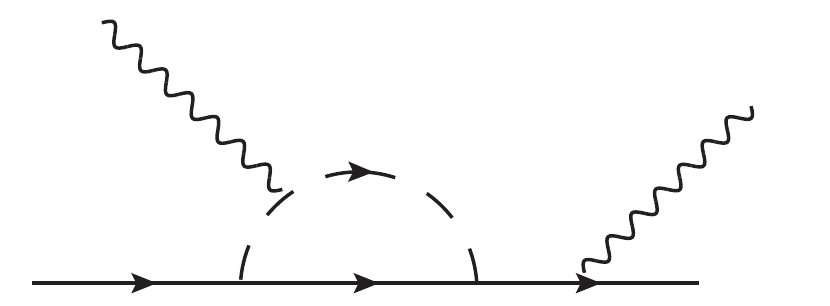}
      \includegraphics[width=0.3\textwidth]{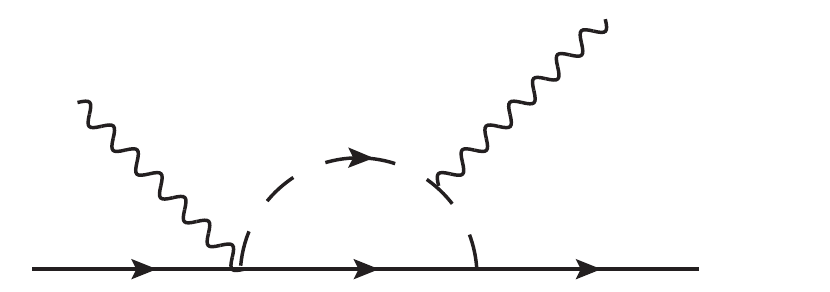}\\
      \includegraphics[width=0.3\textwidth]{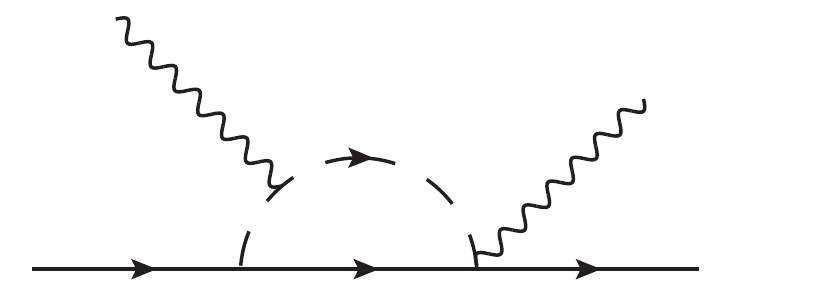}
      \includegraphics[width=0.3\textwidth]{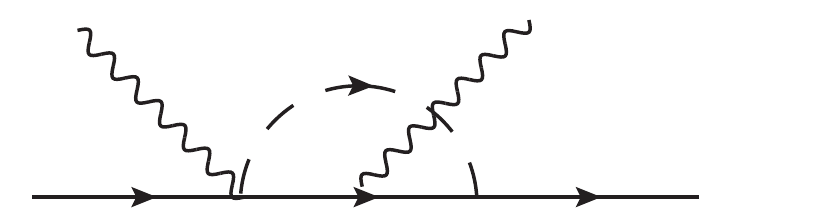}
      \includegraphics[width=0.3\textwidth]{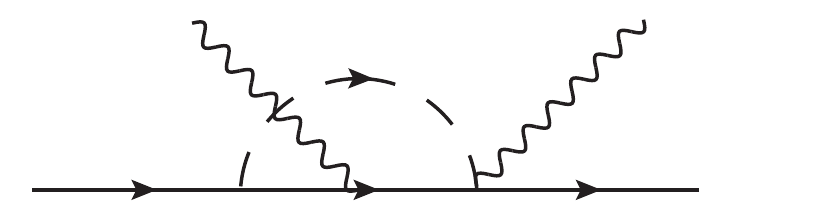}\\
      \includegraphics[width=0.3\textwidth]{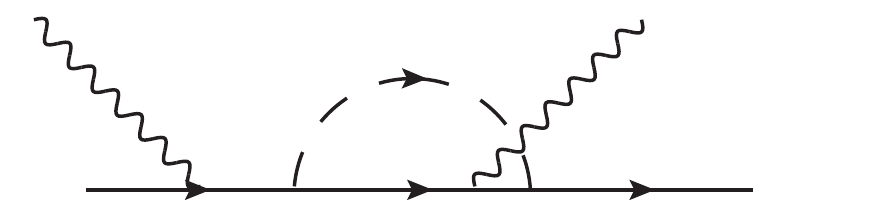}
      \includegraphics[width=0.3\textwidth]{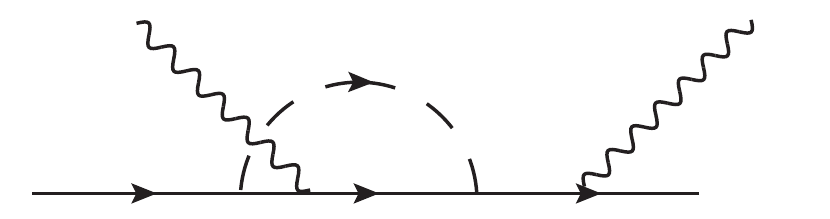}\\
      \includegraphics[width=0.3\textwidth]{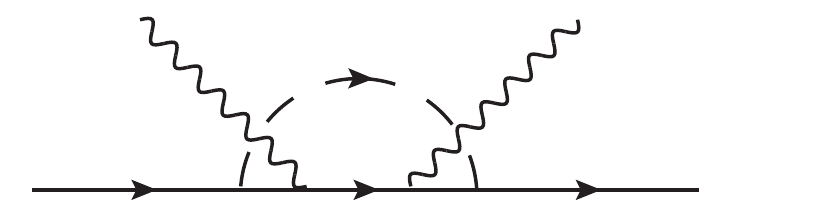}
      \includegraphics[width=0.3\textwidth]{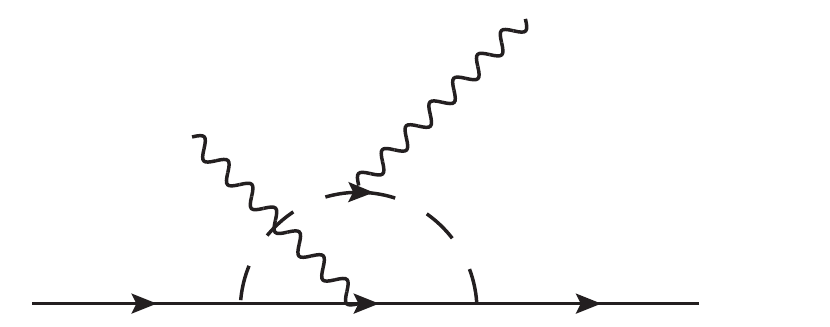}
  \end{center}
  \caption{Diagrams contributing to $\gamma_0$ with isospin-$1/2$ 
    intermediate states. The crossed diagrams are not depicted, but are also calculated.}
  \label{fiso12}
\end{figure}

\begin{figure}
  \begin{center}
    \subfigure[]{
      \label{fdeltree}
      \includegraphics[width=0.3\textwidth]{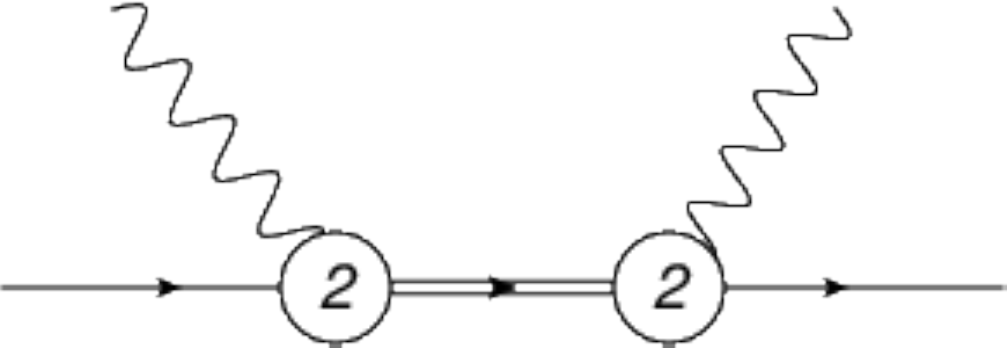}}
    \subfigure[]{
      \label{fdela}
      \includegraphics[width=0.3\textwidth]{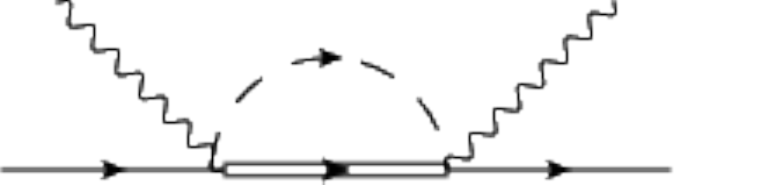}}
    \subfigure[]{
      \label{fdelb}
      \includegraphics[width=0.3\textwidth]{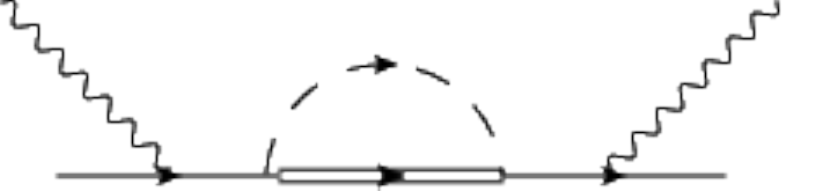}}\\
    \subfigure[]{
      \label{fdelc}
      \includegraphics[width=0.3\textwidth]{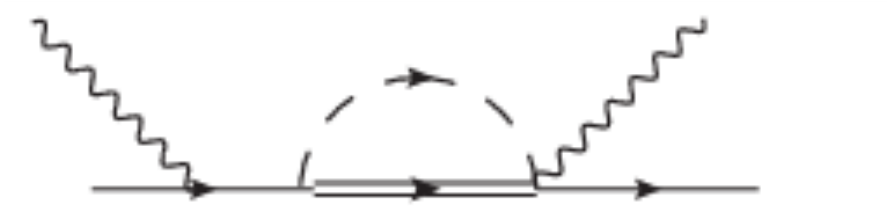}}
    \subfigure[]{
      \label{fdeld}
      \includegraphics[width=0.3\textwidth]{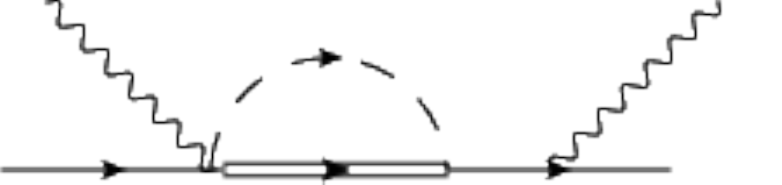}}
    \subfigure[]{
      \label{fdele}
      \includegraphics[width=0.3\textwidth]{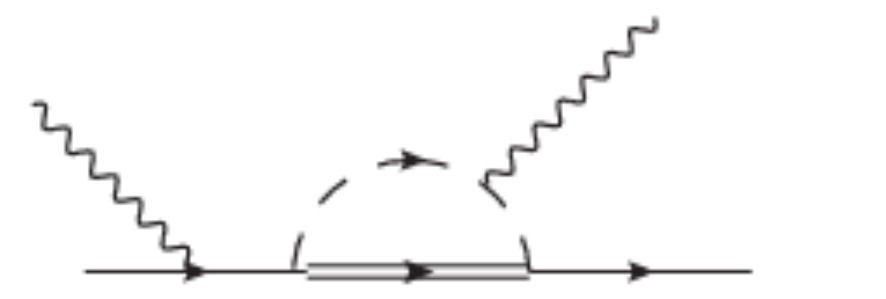}}\\
    \subfigure[]{
      \label{fdelf}
      \includegraphics[width=0.3\textwidth]{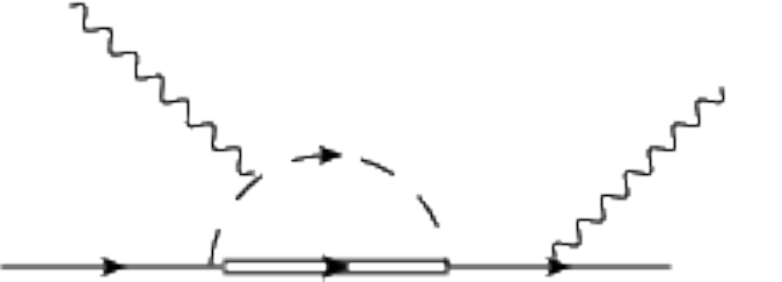}}
    \subfigure[]{
      \label{fdelg}
      \includegraphics[width=0.3\textwidth]{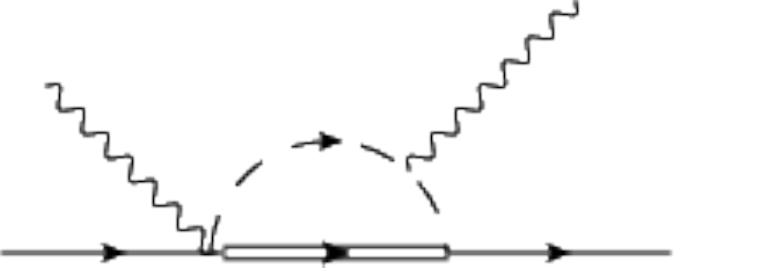}}
    \subfigure[]{
      \label{fdelh}
      \includegraphics[width=0.3\textwidth]{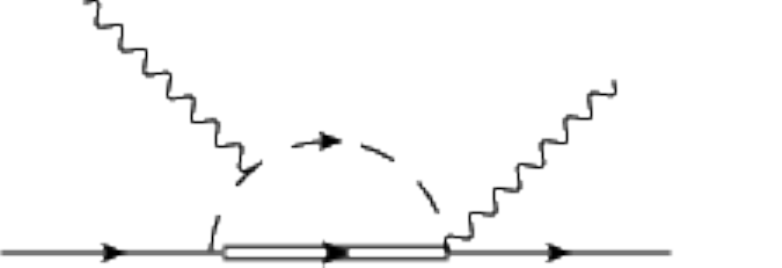}}
  \end{center}
  \caption{Diagrams contributing to $\gamma_0$ with isospin-$3/2$ 
    intermediate states. The crossed diagrams are not depicted, but are also calculated. The one-particle reducible loop diagrams are those from (c) to (g), and the one-particle irreducible ones are (b), (h) and (i).}
  \label{fiso32}
\end{figure}

In Figs.~\ref{fiso12} and~\ref{fiso32} we show all the relevant Compton-scattering diagrams for the extraction of $\gamma_0$ 
up to order $p^{7/2}$.

The extracted numerical results are shown in Table~\ref{tabsu3ours}. First, we show the results in a heavy-baryon ChPT model, as already calculated 
in~\cite{VijayaKumar:2011uw}. This table shows slightly different values, as they were corrected in our work~\cite{Astrid_Diplom}. We then show the 
same calculations done in the fully covariant model. One can immediately see that the values are shifted towards the 
experimental expectation of 
$\left(-1.01\pm0.08\text{(stat)}\pm0.10\text{(syst)}\right)\cdot 10^{-4}$fm$^4$
presented in~\cite{Hildebrandt:2003fm}. But the latter value and also the dispersion relation studies in~\cite{Sandorfi:1994ku}, 
where $\gamma_0^p=-1.34\cdot 10^{-4}$fm$^4$ and $\gamma_0^n=-0.38\cdot 10^{-4}$fm$^4$, 
show that even this shift is not enough to reproduce the experiment. Thus we finally show the results when including the full decuplet degrees of freedom. They reflect the expected behaviour --- a negative polarizability value for both nucleons --- and their size closer to experiment, 
especially when taking into account the error arising from the uncertainty of the constant $g_M$.
Only for $\Sigma^-$ and $\Xi^-$ the uncertainty coming from taking into account the decuplet intermediate states is not significant. The reason for this is that the electromagnetic coupling of these octet baryons to the decuplet is forbidden. Therefore the tree diagrams proportional to $g_M^2$ do not contribute to these two particular states and the decuplet appears only in the form of small contributions from loop diagrams.

\begin{table}[h]
  \begin{center}{\small
    \begin{tabular}{l|*{3}{c}}
      \hline \hline
      \multirow{2}{*}{Baryons}& without decuplet, HBChPT&without decuplet, covariant& with decuplet, covariant\\
	&\cite{VijayaKumar:2011uw,Astrid_Diplom}&\cite{Blin:2015era}&\cite{Blin:2015era}\\
	\hline	$p$&4.69&1.68&-1.64(33)\\
	$n$&4.53&2.33&-1.03(33)\\
	$\Sigma^+$&2.77&0.93&-2.30(33)\\
	$\Sigma^-$&2.54&0.91&0.90\\
	$\Sigma^0$&2.44&1.32&0.47(8)\\
	$\Lambda$&2.62&1.28&-1.25(25)\\
	$\Xi^-$&0.52&0.15&0.13\\
	$\Xi^0$&0.68&0.25&-3.02(33)\\
      \hline \hline
    \end{tabular}}
  \end{center}
  \caption{Numerical values for $\gamma_0$ obtained in our calculations, in units 
    of $10^{-4}$ fm$^4$ in the SU(3) sector.}
  \label{tabsu3ours}
\end{table}

We conclude by stating that the inclusion of the decuplet as an additional degree of freedom to the calculation of baryon polarizabilities is crucial to be able to explain the empirical results. Furthermore, it would be very interesting to study more deeply the $\gamma_0$ values for the hyperons $\Xi^-$ and $\Sigma^-$, e.g. in lattice QCD, as their values are mostly described with spin-1/2 intermediate states, therefore not bringing in the uncertainties of the spin-3/2 sector and serving as a valuable test for the quality of using ChPT methods to describe this quantity.

\section*{Acknowledgements}
This work was supported by the Tomsk State University Competitiveness Improvement Program, by the Russian Federation Program 
"Nauka" (Contract No. 0.1526.2015, 3854),
by the Spanish Ministerio de Econom\'ia y Competitividad and European FEDER funds under Contracts No. FIS2011-28853-C02-01 and FIS2014-51948-C2-2-P, by Generalitat Valenciana under Contract No. PROMETEO/20090090 and by the EU HadronPhysics3 project, Grant Agreement No. 283286. A.N. Hiller Blin acknowledges support from the Santiago Grisol\'ia program of the  Generalitat Valenciana.

\end{document}